\documentclass{eas}
\usepackage{graphicx}
\begin{document}

%%-----------------------------
%%      the top matter
%%-----------------------------
\title{GRB emission in Neutron Star transitions} 
\author{M. A. P\'erez-Garc\'ia}\address{Departament of Fundamental Physics and IUFFyM, University of  Salamanca, Plaza de la Merced s/n 37008 Salamanca, Spain}
\author{F. Daigne}\address{UPMC-CNRS, UMR7095, Institut d'Astrophysique de Paris, 75014 Paris, France}
\author{J. Silk} \address{Oxford Physics, University of Oxford, Keble Road OX1 3RH, Oxford, United Kingdom}
\begin{abstract}
In this contribution we briefly introduce a mechanism for short gamma ray burst emission different from the  usually assumed  compact object binary merger progenitor model.  It is based on the energy release in the central regions of neutron stars. This energy injection may be due to internal self-annihilation of dark matter gravitationally accreted from the galactic halo. We explain how this effect may trigger its full or partial conversion into a quark star  and, in such a case, induce a gamma ray burst with isotropic equivalent energies in agreement with those measured experimentally. Additionally, we show how the ejection of the outer crust in such events may be accelerated enough to produce Lorentz factors over those required for  gamma ray emission. 
\end{abstract}
\maketitle
%%-------------%%-----------------------------
\section{Introduction}
Short gamma ray bursts (SGRB) are highly energetic phenomena in the universe that according to its time duration (\cite{kouveliotou:93}) can be considered to be $\Delta t<2$ s.  Energetically,  they can emit isotropic equivalent energies in the range $E_{\rm iso} \approx 10^{48}-10^{52}$ erg, and some with beamed emission. The beaming factor accounts for the milder emission at the source, $E_{\rm iso}/f_\mathrm{b}$, due to the geometrical finite solid angle, defined as $f_\mathrm{b}=\left(\frac{\Omega}{4\pi}\right)^{-1}$. Typically, there is a large spread in $f_\mathrm{b}$ but it can be $\sim 10-100$. To date, although higly uncertain, such a rapid release of huge energies is thought to correspond to the merger of two compact objects, each of them possibly being either a neutron star (NS) or a black hole (BH).
NSs are compact-sized objects with a typical mass $M\sim1.5 M_{\odot}$ and radius $R\sim12$ km born in a supernova event. The interior of these objects is largely unknown and periodically revisited (\cite{glen}) but the large densities suppossedly attained in the core of these objects of about $\rho\sim 4\,10^{14} \rm g/cm^3$ well overpass the nuclear saturation density of regular finite nuclei. In this context, matter is well described by a degenerate system of nucleons where temperature effects are negligible as compared to the Fermi energies of the nuclear species population, $k_BT/E_F<<1$. It is expected that the $\it{core}$, most of the NS, is  formed by a dense soup of nucleons or even heavier baryons bearing strangeness. The equation of state (EoS) describing the interior is largely uncertain but mainly concerns weak and strong interactions and can be treated using a variety of methods. We will consider in this contribution a relativistic field model with two parameterizations, the TM1 (\cite{eos1}) and TMA (\cite{eos2}) EoSs. 

These objects have a strong gravitational potential as probed, for example, in the absorption lines in the spectra of 28 bursts of the low-mass X-ray binary  EXO 0748-676 (\cite{cottam}). The mass-to-radius ratio was probed in this way, since $z=(1-R_S/R)^{-1/2} -1$ with $R_S=2GM/c^2$ the Schwarzschild radius, and it was possible to determine that $z=0.35$ for this object, a huge value as compared to cluster masses of $\sim10^{14}M_\odot$, where it is estimated to be $z\sim10^{-5}$ or the sun $z\sim 10^{-6}$ (\cite{lopresto}). This provided the first observational direct evidence that NSs are indeed made of tightly packed matter and confirms that could be considered very effective accretors from a companion or from matter distributed in the galactic halo. 

In this light the sun has been considered of interest for the indirect detection of dark matter (DM). This so-far undetected component of our universe is thought to amount  $\sim23\%$, for a review see (\cite{bertone}).
There is a number of DM particle candidates in current theoretical models so called {\it beyond} the Standard model where they naturally arise, the most popular being the lightest supersymmetric particle, the neutralino. As it seems there are some indications that either from accelerator, direct or indirect  searches the discovery may be not too far. Particle DM candidates with masses $m_{\chi}\sim4-12$ $\rm GeV/c^2$ in the direct and $m_{\chi}\sim130$ $\rm GeV/c^2$ in the indirect  search are currently under debate. In this sense the fact that DM could be a self-annihilating Majorana fermion could lead to dramatic consequences from the astrophysical point of view as we will explain in what follows. 

\section{SGRB engine model in brief}
It has been claimed that NSs could accrete DM from the galactic halo and due to a density enhancement at the core of the NS, spark seeding based on DM self annihilation may be possible (\cite{perez:10}). This sort of {\it Trojan mechanism}  would allow, in principle, the release of energies comparable to binding of quarks in baryons and, therefore, the deconfinement of the quark content. Once quark lumps  are formed they are energetically very stable and they could grow or coalescence helped by the pressure softening of this type of matter. It has been studied (\cite{burning}) that a macroscopic conversion burning front may indeed be possible (partially) fully converting the NS into a (hybrid) quark star (QS) (\cite{perez:11}).
Typically, the object converting has an energy reservoir that is a factor of order unity  the gravitational binding energy, and if such a transition takes place the mass of the resultant QS is close to the initial NS $(M_{QS} \approx M_{NS})$. When it is converted a fraction of the two object binding energies  $f_\mathrm{QS/NS} \approx \left|\left(\frac{M_\mathrm{QS}}{M_\mathrm{NS}}\right)^2\left(\frac{R_\mathrm{NS}}{R_\mathrm{QS}}\right)-1 \right|$ is released. In order to see if this type of single non-repeating catastrophic events have an occurrence rate compatible with current limits on those of SGRBs we compare them to those of Supernova type II formation and we see (\cite{perez:12}) that even considering the large uncertainty in the values of beaming factors, indeed there must be a non-trivial  time delay probability distribution and only in rare cases such events must take place since $\frac{\mathcal{R}_\mathrm{SGRB}}{\mathcal{R}_\mathrm{NS\to QS,max}^\mathrm{(SNII)}} \simeq \left( 8\times 10^{-4} \to 3\times 10^{-3}\right) \left(\frac{\left\langle f_\mathrm{b}\right\rangle}{50}\right)$.

Regarding the properties in the host galaxy there is no definite place for the SGRBs, being allowed in all regions in either type of galaxy. There is however less correlation than with respect to long GRBs that show a correspondance with central regions of star forming galaxies. In this model the progenitors (i.e. NSs) show a natural mechanism to un-correlate due to a natal kick velocity of the order $v\sim10^3$ $\rm km/s$ providing off-sets of the order $d\sim v \tau$ where $\tau$ is the delay to the conversion. In this model it is highly uncertain since it depends on the drift history of the NS and the inhomogeneous cluster of the DM environment traversed over time scales of $\tau \sim10^3-10^7$ yr according to the details of the DM steady accretion and deduced ages of {\it regular} NSs.

As for the energetics, the amount of kinetic energy of the expelled outer crust is a fraction of the gravitational energy $E_\mathrm{kin}\simeq f_\mathrm{ej} \Delta E_\mathrm{grav}$ if acceleration is complete.
As explained, the isotropic equivalent energy  that could be radiated as gamma-rays by such an ejecta can be estimated by 
\begin{equation}
E_\mathrm{\gamma,iso}\simeq 3.5\times 10^{51}\, \left(\frac{f_\mathrm{b}}{100}\right) \left(\frac{f_\mathrm{\gamma}}{0.1}\right)\left(\frac{f_\mathrm{ej}}{10^{-3}}\right)\left(\frac{R_{\rm NS}}{12 \, \rm km}\right)^{-1}\left(\frac{M}{1.5\, M_{\odot}}\right)^2 \, \mathrm{erg}
\, ,
\label{eq:egammiso}
\end{equation}
where $f_\mathrm{\gamma}$ is the efficiency of gamma-ray energy extraction from the ejecta and could range from $\sim 0.01-0.1$. This estimate of $E_\mathrm{\gamma,iso}$ is in reasonable agreement with observations of SGRBs: 
the NS$\to$QS conversion scenario investigated here can reproduce measured energies in SGRBs  for $f_\mathrm{b} f_\mathrm{\gamma} f_\mathrm{ej} \simeq 3\times 10^{-4}-0.3$. 
%%%%%%%%%%%%%%%%%%%%%%%%%%%%%%%%%%%%%%%%%%%%%%%%%%%%%%%%%%%%%%%%%
\begin{figure}%[hbtp]
\begin{center}
\includegraphics [angle=-90,scale=0.5] {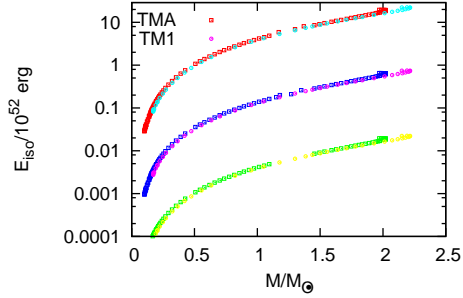}
\caption{Isotropic equivalent energy as a function of the progenitor NS mass (in $M_{\odot}$ units) as obtained with the TM1 (squares) and TMA (circles) EoS. From top to bottom a value of $f_\mathrm{b} f_\mathrm{\gamma} f_\mathrm{ej} = 0.3, 0.01, 3\times 10^{-4}$ has been assumed.
}
\label{fig:1}
\end{center}
\end{figure}
%%%%%%%%%%%%%%%%%%%%%%%%%%%%%%%%%%%%%%%%%%%%%

In  Figure~\ref{fig:1} we show the isotropic equivalent energy (logarithmic scale) as a function of the progenitor NS mass, obtained with the TM1 (squares) and TMA (circles) EoS. From top to bottom a value of the product of three efficiency fractions, $f_\mathrm{b} f_\mathrm{\gamma} f_\mathrm{ej} \simeq 0.3, 0.01, 3\times 10^{-4}$ has been assumed. We can see that the major dependence is due to the microphysics efficiency of the model and ejection but the interior EoS mainly affects the possibility of more massive progenitors. This is quite natural consequence of the fact that the mass-radius relationships for both EoS are similar in shape but they attain different maximum masses. Due to the fact that typical energies are in the range $E_{\gamma,\rm iso} \approx 10^{48}-10^{52}$ erg, it leads to consider that smaller efficiency fractions are even able to produce SGRBs in light of this mechanism. However further details and modelling is needed.

As for Lorentz factors they can be estimated from the ejected mass, $M_\mathrm{ej}$, and  $E_{\mathrm{kin}}$ as,
\begin{equation}
\Gamma_\mathrm{max}=\frac{E_\mathrm{kin}}{M_\mathrm{ej} c^2} 
\simeq 19\, \left(\frac{f_\mathrm{ej}}{10^{-3}}\right)\left(\frac{R_{\rm NS}}{12 \, \rm km}\right)^{-1}\left(\frac{M}{1.5\, M_{\odot}}\right)^2 \left(\frac{M_\mathrm{ej}}{10^{-5}\, M_\odot}\right)^{-1} \, .
\end{equation}
 As the fraction of the energy injected in the outer crust is not too small ($f_\mathrm{ej} \geq 10^{-3}$) $\Gamma_\mathrm{max}>15$, in agreement with the observational constraints and even ultra-high relativistic ejecta with $\Gamma_\mathrm{max}>100$ are allowed if only the {\it outer} crust is expelled. The mass in the outer crust can be computed from the integration of the structure  equations (\cite{tov}) (for the non-rotating case) from the neutron drip,  $\rho_\mathrm{ND} \approx 4\,10^{11}$ $ \rm g/cm^3$, out to the very external radius in the NS.

%%%%%%%%%%%%%%%%%%%%%%%%%%%%%%%%%%%%%%%%%%%%%%%%%%%%%%%%%%%%%%%%%
\begin{figure}%[hbtp]
\begin{center}
\includegraphics [angle=-90,scale=0.5] {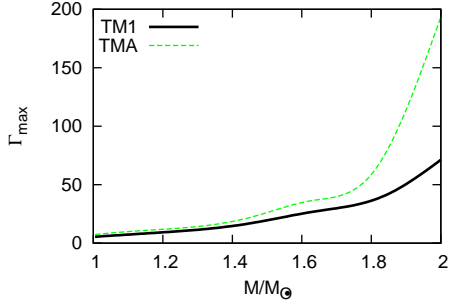}
\caption{Maximum Lorentz factor as a function of the mass (in $M_{\odot}$ units) of the progenitor NS as computed with the TM1  (solid line) and TMA (dashed line) EoS. An ejection fraction of $f_{\rm ej}=10^{-2}$ has been assumed.
}
\label{fig:2}
\end{center}
\end{figure}
%%%%%%%%%%%%%%%%%%%%%%%%%%%%%%%%%%%%%%%%%%%%%

In  Figure~\ref{fig:2} we show the maximum Lorentz factor $\Gamma_\mathrm{max}$ as a function of the NS mass (in $M_{\odot}$ units) for the TM1 (solid line) and TMA (dashed line) EoSs. For each value of the progenitor mass, the outer crust value is obtained as the mass from the radial value where the baryonic density falls below the neutron drip density. An ejection fraction of $f_{\rm ej}=10^{-2}$ has been assumed as an average typical value. We can see that for  masses  above $M\simeq 1.4 M_{\odot}$ gamma ray emission is allowed, and typical NS meassured masses confirm this range.
In particular, the value of  $1.97\pm0.04\, M_{\odot}$ for the mass of PSR J1614-223048 measured  using Shapiro time delay (\cite{demorest}) can be reached in this context using the TMA EoS description. In addition to the gamma ray emission it is expected a multi-messenger correlated signal emitted in gravitational waves and neutrinos. In this sense advanced versions of LIGO/VIRGO and KM3 net detectors could detect this in the coming future.

\section{Conclusions}
In the present contribution we discuss  whether a non-repeating cataclysmic event of the type NS$\to$QS transition driven by DM could produce SGRBs and what would be  typical values of isotropic energies and maximum Lorentz factors. We compare the energy released by SGRBs with the energetics of the progenitor model that we propose for two popular NS EoS. We discuss the short GRB event rate in light of this scenario as compared to observed rates, as well as the natural delay time between the regular NS phase and the QS formation and the properties of the host galaxies of
short GRBs. We analyze crust masses for the same EoSs that could
be expelled due to the NS$\to$QS conversion and we obtain that Lorentz factors could be ultra-relativistic for large enough progenitors.

We thank the MICINN (Spain) MULTIDARK, FIS-2009-07238 and FIS2011-14759-E, FIS2012-30926 projects and the  ESF-funded COMPSTAR project for partial financial support. MAPG acknowledges IAP for its kind hospitality while this work was completed.
%%-------------------%%-----------------------------


\begin{thebibliography}{99}

\bibitem[{Bhattacharyya et al.\, 2006}]{burning} Bhattacharyya A. et al. 2006, Physical Review C 74, 065804 

\bibitem[{Bertone\, 2010}]{bertone} Bertone G. , ed., Particle Dark Matter: Observations, Models and Searches. Cambridge University Press, 2010, ISBN 978-0-521-76368-4.

\bibitem[Cottam et al.\,2002]{cottam} Cottam J., Paerels F. and Mendez M., 2002, Nature, 420 51.


\bibitem[{Demorest et al.\, 2010}]{demorest}  Demorest P. B. et al., 2010, Nature, 467, 1081.

\bibitem[{Glendenning\, 2000}]{glen} Glendenning N.K. 2000, {\it Compact stars}, Ed. Springer-Verlag, New York.

\bibitem[ Kouveliotou et al.\, 1993]{kouveliotou:93} Kouveliotou, C. \etal\ 1993, ApJL, 413, L101.

\bibitem[Lopresto et al.\, 1991]{lopresto} Lopresto, J. C., Schrader, C. and Pierce, A. K. 1991, ApJ 376, 757.

\bibitem[{Oppenheimer and Volkoff\,1939}]{tov}  Oppenheimer J. R., Volkoff G. M., 1939, Phys. Rev. 55, 374.

\bibitem[Perez-Garcia et al.\, 2010]{perez:10} 
{Perez-Garcia, M.~A., Silk, J., \& Stone, J.~R.\ 2010, Physical Review Letters, 105, 141101.}

\bibitem[{Perez-Garcia and Silk\, 2012}]{perez:11} Perez-Garcia, M. A., Silk, J. 2012,  Phys. Lett. B, 711, 6.

\bibitem[{Perez-Garcia et al.\, 2012}]{perez:12} Perez-Garcia, M. A., Daigne, F. and Silk, J., 2012, submitted.

\bibitem[{Sugahara et al.\,1994}]{eos1} Sugahara Y. and Toki H., 1994,  Nucl. Phys. A 579, 557.

\bibitem[{Toki et al.\,1995}]{eos2} Toki H. et al., 1995, Nuclear Physics A 588, 357 .

\end{thebibliography}
\end{document}